\documentclass{ws-procs975x65}

\begin{document}



\title{PAIR WINDS IN SCHWARZSCHILD SPACETIME WITH APPLICATION TO
STRANGE STARS}

\author{A.G. AKSENOV}

\address{Institute of Theoretical and Experimental Physics,\\
B.~Cheremushkinskaya, 25, Moscow 117218, Russia \\
E-mail: alexei.aksenov@itep.ru}

\author{M. MILGROM and V.V. USOV}

\address{Center for Astrophysics, Weizmann Institute,\\
Rehovot 76100, Israel \\
E-mail: moti.milgrom@weizmann.ac.il\\
vladimir.usov@weizmann.ac.il}


\begin{abstract}
We present the results of numerical simulations of stationary,
spherically outflowing, $e^\pm$ pair winds, with total luminosities
in the range $10^{34}- 10^{42}$ ergs~s$^{-1}$. In the concrete
example described here, the wind injection source is a hot, bare,
strange star, predicted to be a powerful source of $e^\pm$ pairs
created by the Coulomb barrier at the quark surface. We find that
photons dominate in the emerging emission, and the emerging photon
spectrum is rather hard and differs substantially from the thermal
spectrum expected from a neutron star with the same luminosity. This
might help distinguish the putative bare strange stars from neutron
stars.
\end{abstract}

\bodymatter

\section{Introduction}
\label{intro}

For an electron-positron ($e^\pm$) wind out-flowing spherically from
a surface of radius $R$ there is a maximum pair luminosity,
$L_\pm^{\rm max} = {4\pi m_ec^3R\Gamma^2/\sigma_{\rm T}}\simeq
10^{36}(R/10^6{\rm cm})\Gamma^2\,{\rm ergs~s}^{-1}$, beyond which
the pairs annihilate significantly before they escape, where
$\Gamma$ is the pair bulk Lorentz factor, and $\sigma_{\rm T}$ the
Thomson cross section. Recently we developed a numerical code for
solving the relativistic kinetic Boltzmann equations for pairs and
photons in Schwarzschild geometry. Using this we considered a
spherically out-flowing, non-relativistic ($\Gamma\sim 1$) pair
winds with injected pair luminosity $\tilde L_\pm$ in the range
$10^{34}-10^{42}$ ergs~s$^{-1}$, that is $\sim
(10^{-2}-10^{6})L_\pm^{\rm max}$ (Aksenov et al. 2003, 2004, 2005).
While our numerical code can be more generally employed, the results
presented in this paper are for a hot, bare, strange star as the
wind injection source. Such stars are thought to be powerful (up to
$\sim 10^{51}$ ergs~s$^{-1}$) sources of pairs created by the
Coulomb barrier at the quark surface (Usov 1998, 2001).

\section{Formulation of the problem}
We consider an $e^\pm$ pair wind that flows away from a hot, bare,
unmagnetized, non-rotating, strange star. Space-time outside the
star is described by  Schwarzschild's metric with the line element
\begin{equation}
  ds^2
 =-e^{2\phi}c^2
dt^2+e^{-2\phi}dr^2+r^2(d\vartheta^2+\sin^2\vartheta\,
d\varphi^2)\,, \label{ds1}
\end{equation}
where $e^{\phi}=\left(1-{r_g/ r}\right)^{1/2}$ and $r_g={2GM/
c^2}\simeq 2.95 \times 10^5 ({M/ M_\odot})$ cm.

We use the general relativistic Boltzmann equations for the pairs
and photons, whereby the distribution function for the particles of
type $i$, $f_i(p, \mu, r, t)$, satisfies
\begin{eqnarray}
  \frac{e^{-\phi}}{c}\frac{\partial f_i}{\partial t}
 +\frac{1}{r^2}\frac{\partial}{\partial r}(r^2\mu e^{\phi}\beta_i
f_i)
 -\frac{e^{\phi}}{p^2}\frac{\partial}{\partial p}
  \left(
    p^3 \mu \frac{\phi'}{\beta_i} f_i
  \right) \nonumber \\
 -\frac{\partial}{\partial\mu}
  \left[
    (1-\mu^2)e^\phi
          \left(\frac{\phi'}{\beta_i}-\frac{\beta_i}{r}\right) f_i
  \right]
 =\sum_q(\bar\eta^q_i-\chi^q_i f_i).
\label{dfi}
\end{eqnarray}
Here, $\mu$ is the cosine of the angle between the radius and the
particle momentum ${\bf p}$, $p=|{\bf p}|$, $\beta_e=v_e/c$,
$\beta_\gamma =1$, and $v_e$ is the velocity of electrons and
positrons. Also, $\bar\eta_i^q$ is the emission coefficient for the
production of a particle of type $i$ via the physical process
labelled by $q$, and $\chi_i^q$ is the corresponding absorption
coefficient. The processes we include are listed in the following
Table.

\begin{table}
\caption{Physical Processes Included in Simulations}
\begin{center}
\begin{tabular}{ll}
  \hline \hline
  Basic Two-Body & Radiative \\
  Interaction & Variant \\
  \\ \hline
  M{\o}ller and Bhaba &  \\
  scattering & Bremsstrahlung \\
  $ee\rightarrow ee$ & $ee\leftrightarrow ee\gamma$ \\ \hline
  Compton scattering \,\,\,\,\,\,& Double Compton scattering \\
  $\gamma e\rightarrow \gamma e$ & $\gamma e\leftrightarrow \gamma
e\gamma$
  \\ \hline
  Pair annihilation & Three photon annihilation \\
  $e^+e^-\rightarrow \gamma\gamma$ & $e^+e^-\leftrightarrow
\gamma\gamma\gamma$
  \\ \hline
  Photon-photon &  \\
  pair production &   \\
  $\gamma\gamma\rightarrow e^+e^-$ &   \\ \hline
\end{tabular}
\end{center}
\end{table}

\section{Numerical results}
For injected pair luminosity $\tilde L_\pm$ higher than $\sim
10^{34}$ ergs~s$^{-1}$, the emerging emission consists mostly of
photons (see Fig.~1, left panel). This simply reflects the fact that
in this case the pair annihilation time $t_{\rm ann}\sim (n_e
\sigma_{\rm T}c)^{-1}$ is less than the escape time $t_{\rm esc}\sim
R/c$. There is an upper limit to the rate of emerging pairs $\dot
N_e^{\rm max}\simeq 10^{43}$~s$^{-1}$ (see Fig.~1, right panel).

As $\tilde L_\pm$ increases from $\sim 10^{34}$ to $10^{42}$
ergs~s$^{-1}$, the mean energy of emergent photons decreases from
$\sim 400$ keV to 40 keV, as the spectrum changes in shape from that
of a wide annihilation line to nearly a blackbody spectrum with a
high energy ($> 100$ keV) tail (see Fig.2).

\begin{figure}
\centering \resizebox{1\textwidth}{!}{
  \includegraphics[25,14][272,200]{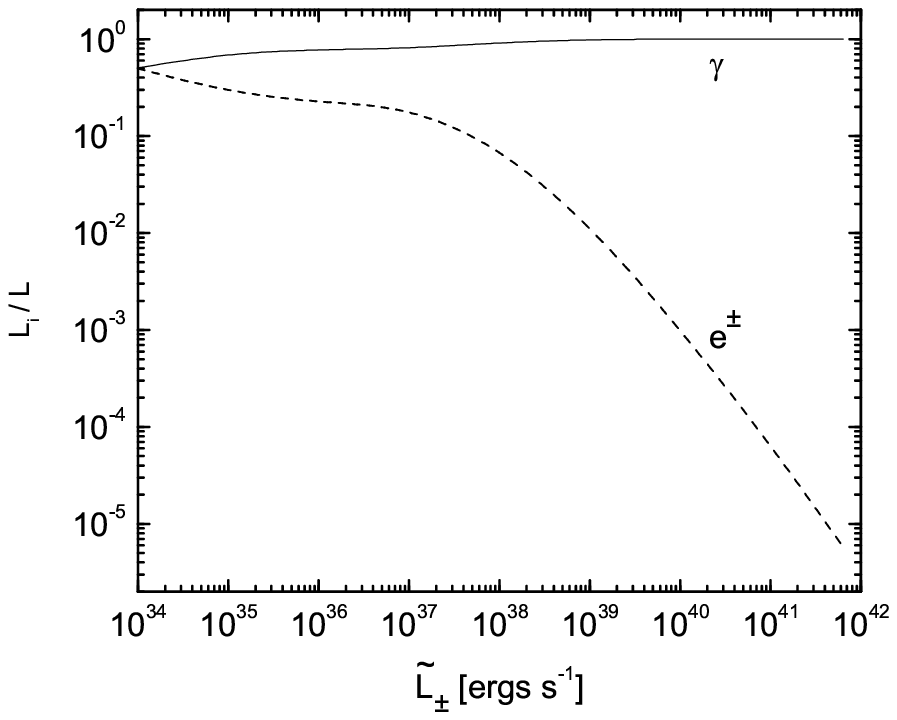}
  \resizebox{0.74\textwidth}{!}{\includegraphics[14,12][282,200]{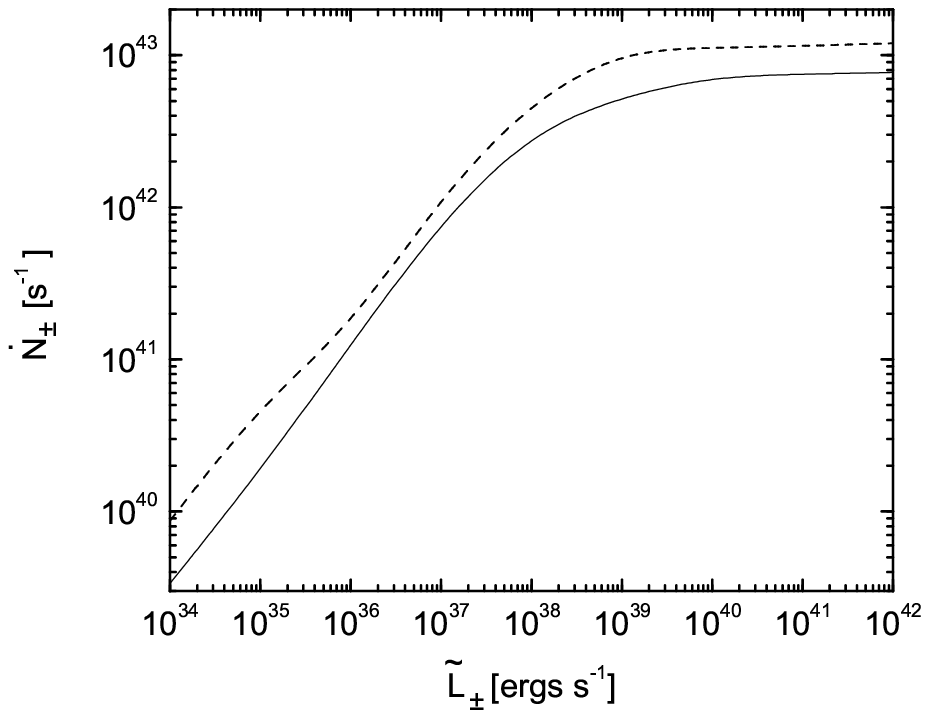}}
}
\caption{LEFT: The fractional emerging luminosities in pairs (dashed
line) and photons (solid line) as functions of the injected pair
luminosity, $\tilde L_\pm$. RIGHT: Number rate of emerging pairs as
functions of $\tilde L_\pm$ (solid line). The case where gravity has
been neglected is shown by the dashed line.}
\label{Aksenovf3.eps}       
\end{figure}

\begin{figure}
\centering
\resizebox{1.01\textwidth}{!}{
  \includegraphics[25,14][290,200]{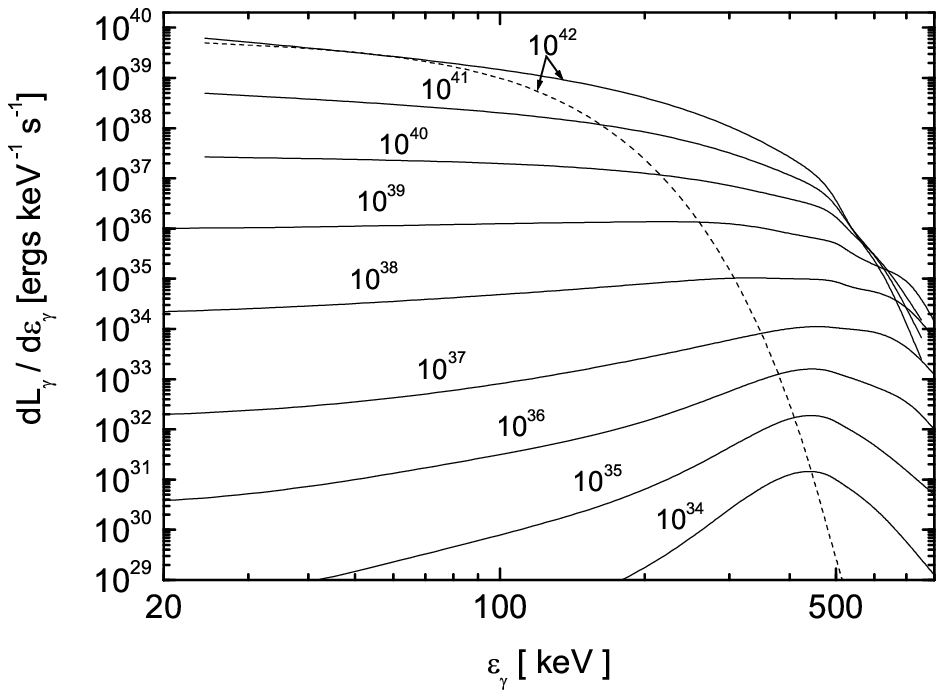}
  \resizebox{0.72\textwidth}{!}{\includegraphics[14,13][282,200]{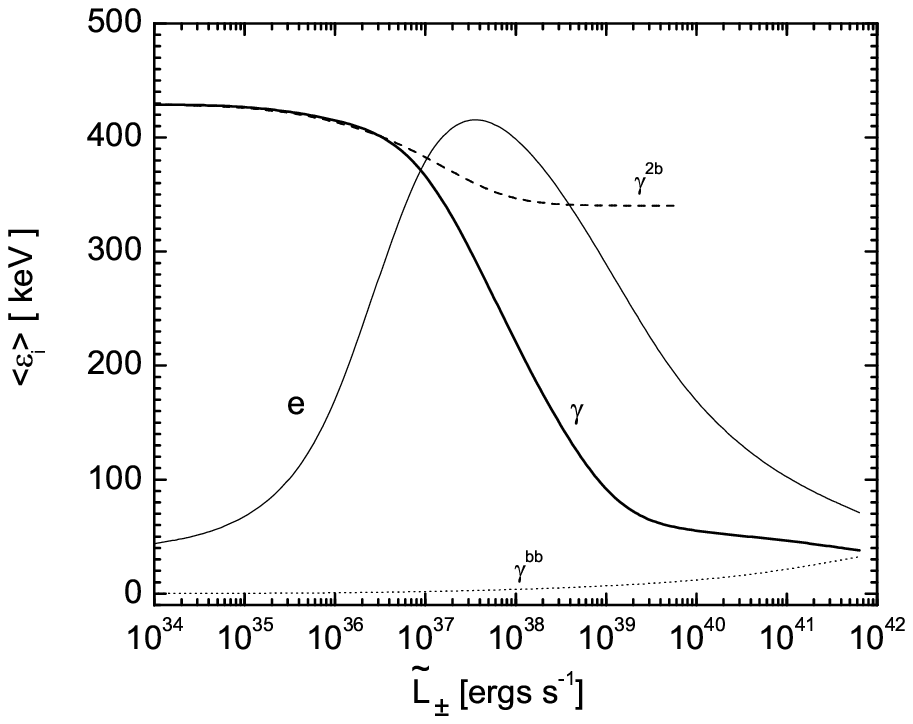}}
}
\caption{LEFT: The energy spectrum of emerging photons for different
values of $\tilde L_\pm$, as marked on the curves. The dashed line
is the spectrum of blackbody emission. RIGHT: The mean energy of the
emerging photons (thick solid line) and electrons (thin solid line)
as a function of $\tilde L_\pm$. For comparison, we show as the
dotted line the mean energy of blackbody photons for the same energy
density as that of the photons at the photosphere. Also shown as the
dashed line is the mean energy of the emerging photons in the case
when only two particle processes are taken into account.}
\label{Aksenovf3.eps}       
\end{figure}

\section*{Acknowledgments}
The research was supported by the Israel Science Foundation of the
Israel Academy of Sciences and Humanities.

\end{document}